\documentclass[12pt]{article}
\usepackage{graphicx}
\csname @addtoreset\endcsname{equation}{section}

\def\bseq{\begin{subequation}}  % = 1a 1b
\def\eseq{\end{subequation}}
\def\bsea{\begin{subeqnarray}}  % = 1.1a 1.1b
\def\esea{\end{subeqnarray}}
\newcommand{\bbox}{\lower.2ex\hbox{$\Box$}}

\newcommand{\beq}{\begin{equation}}
\newcommand{\eeq}{\end{equation}}
\newcommand{\bea}{\begin{eqnarray}}
\newcommand{\eea}{\end{eqnarray}}
\newcommand{\ena}{\end{eqnarray}}

\renewcommand{\d}{\delta}
\renewcommand{\th}{\theta}

\newcommand{\pa}{\partial}
\newcommand{\pab}{{\bar{\pa}}}
\newcommand{\g}{\gamma}

\newcommand{\e}{\epsilon}

\renewcommand{\l}{\lambda}

\newcommand{\m}{\mu}

\newcommand{\n}{\nu}

\newcommand{\s}{\sigma}

\begin{document}

\begin{titlepage}
{\hbox to\hsize{October  2003 \hfill
{Bicocca--FT--03--30}}}
{\hbox to\hsize{${~}$ \hfill
{FNT/T--2003/11}}}
\begin{center}
\vglue .06in
\vskip 15pt
{\Large\bf  Some properties of the integrable noncommutative  
sine--Gordon system}
\\[.35in]
Marcus T. Grisaru\footnote{grisaru@physics.mcgill.ca}\\
{\it Physics Department, McGill University \\
Montreal, QC Canada H3A 2T8 }
\\
[.2in]
Liuba Mazzanti\footnote{liuba@pcteor1.mi.infn.it}\\
{\it Dipartimento di Fisica, Universit\`a degli studi di Milano-Bicocca\\
piazza della Scienza 3, I-20126 Milano, Italy}
\\
[.2in]
Silvia Penati\footnote{silvia.penati@mib.infn.it}\\
{\it Dipartimento di Fisica, Universit\`a degli studi di
Milano-Bicocca\\ 
and INFN, Sezione di Milano, piazza della Scienza 3, I-20126 Milano, 
Italy}
\\
[.2in]
Laura Tamassia\footnote{laura.tamassia@pv.infn.it}\\
{\it Dipartimento di Fisica Nucleare e Teorica, Universit\`a degli studi di 
Pavia and INFN, Sezione di Pavia, via Ugo Bassi 6, I-27100 Pavia, Italy}
\\[.3in]

{\bf ABSTRACT}\\[.0015in]
\end{center}
In this paper we continue the program, initiated in Ref. \cite{us}, 
to investigate an integrable noncommutative version of the sine--Gordon model. 
We discuss the origin of the extra constraint which the field function has
to satisfy in order to guarantee classical integrability. We show that 
the system of constraint plus dynamical equation of motion 
can be obtained by a suitable reduction of a noncommutative version of 
4d self--dual Yang--Mills theory. 
The field equations can be derived from an action which is the sum of two 
WZNW actions with cosine potentials corresponding to a complexified 
noncommutative $U(1)$ gauge group.
A brief discussion of the relation with the bosonized noncommutative 
Thirring model is given. In spite of integrability we show that the
S-matrix is acausal and particle production takes place.

\vskip 5pt
${~~~}$ \newline

\end{titlepage}

\section{Introduction}

Field theories defined in a noncommutative (NC) geometry have been recently
the subject of detailed investigations, partly because noncommutativity 
emerges naturally in the low--energy dynamics of branes in a constant B field 
\cite{strings},  but also because they have interesting features
of their own. 

In this context, an interesting question is how noncommutativity could 
affect the dynamics of exactly solvable field theories, as for instance 
two--dimensional integrable theories. A common feature of these systems is 
that the existence of an infinite chain of {\em local} conserved currents
is guaranteed by the fact that the equations of motion can be 
written as zero curvature conditions for a suitable set of covariant 
derivatives \cite{curvature}. In some cases, as for example the ordinary 
sine--Gordon or sigma models, an action is also known which generates
the integrable equations according to an action principle. 

Noncommutative versions of these models are intuitively defined as models 
which reduce to the ordinary ones when the noncommutation parameter 
$\theta$ is removed.
In general these generalizations are not unique as one can construct 
different NC equations of motion which collapse to the same expression
when $\theta$ goes to zero   
\footnote{As a simple example, we can
consider a scalar free field theory where the ordinary equations of motion
can have different NC counterparts, as for instance the equation itself or
$\pa_\mu ( e_{\ast}^{-i\phi} \ast \pa^\mu e_{\ast}^{i\phi}) =0$.}.

For two dimensional integrable systems, a general criterion
to restrict the number of possible NC versions is to require the classical
integrability to survive in NC geometry. This suggests that any NC 
generalization should be performed at the level of equations of motion by
promoting the standard zero curvature techniques. This program has been 
worked out 
for a number of known integrable equations in Refs. \cite{NCEOM,NCbur}.

In the case of the sine--Gordon system 
\beq
S = \frac{1}{\pi\l^2} \int d^2z \left[ \pa \phi \pab \phi -
2 \g ( \cos{\phi} - 1) \right]
\label{oaction}
\eeq
we might define the ``natural'' NC generalization as the 
model obtained by promoting the ordinary product to the $\ast$--product in the
action (\ref{oaction}) and consequently in the equations of motion
which become $\pa \pab \phi = \g \sin_{\ast}{\phi}$.
However, this model seems to be affected by some 
problems both at the classical and the quantum level.
 
At the classical level it does not seem to be integrable 
since the ordinary currents promoted to NC currents by replacing the products
with $\ast$--products are not conserved \cite{us}. Moreover, 
we don't know how to find a systematic procedure to construct
conserved currents since the equations of motion cannot be obtained as
zero curvature conditions (a discussion about the lack of integrability for
this system is also given in Ref. \cite{CM}).

At the quantum level the renormalizability properties of the ordinary
model (\ref{oaction}) defined for $\l^2 < 4$ seem to be destroyed by NC. 
The reason is quite simple 
and can be understood by analyzing the structure of the divergences 
of the NC model compared to the ordinary ones \cite{coleman, amit}.
In the $\l^2 < 4$ regime the only divergences come from multitadpole diagrams.
In the ordinary case the $n$--loop diagram gives a contribution 
$(\log{m^2a^2})^n$ where $a$ and $m$ are the UV and IR cut--offs respectively.
This result is independent of the number $k$ of external fields and of
the external momenta. As a consequence the
total contribution at this order can be resummed as
$\g (\log{m^2a^2})^n ( \cos{\phi} - 1)$ and the divergence is cancelled
by renormalizing the coupling $\g$. This holds at any order $n$ and the 
model is renormalizable. 

In the NC case the generic vertex from the expansion of $\cos_{\ast}{\phi}$
brings nontrivial phase factors which depend on the momenta 
coming out of the vertex and on the NC parameter. 
The final configuration of phase factors associated
to a given diagram depends on the order we use to contract the fields 
in the vertex. Therefore, in the NC case the ordinary $n$--loop diagram 
splits into a planar and a certain number of nonplanar configurations, where
the planar one has a trivial phase factor whereas the nonplanar diagrams
differ by the configuration of the phases (for a general discussion see 
Refs. \cite{filk, tadpole}).
The most general NC multitadpole diagram is built up by combining planar 
parts with nonplanar ones where two or
more tadpoles are intertwined among themselves or with external legs. 
Since the nonplanar subdiagrams are convergent \cite{tadpole, liuba} a 
generic  $n$--loop diagram contributes to the divergences of the theory 
only if it contains a nontrivial planar 
subdiagram. However, different $n$--loop diagrams with different 
configurations of planar and nonplanar parts give divergent contributions
whose coefficients depend on the number $k$ of external fields and
on the external momenta.  
A resummation of the divergences to produce a cosine potential is not 
possible anymore and the renormalization of the couplings of the model is 
not sufficient to make the theory finite at any order. Noncommutativity 
seems to deform the cosine potential at the quantum level and the theory 
loses the renormalizability properties of the corresponding commutative model.

Thus, the ``natural'' generalization of sine--Gordon is not satisfactory 
and one must
look for a different NC generalization compatible with integrability 
and/or renormalizability.

In Ref. \cite{us} we considered the problem of classical integrability.  
The method of the {\em bicomplex} implemented in NC
geometry was used to construct a NC version of the sine--Gordon equation 
which is integrable, and the first few nontrivial currents were constructed
and studied. The equations of motion for that system are quite unexpected and
do not resemble the ones of the ``natural'' generalization. Precisely,
what emerges in the NC case is a system of two coupled equations of motion
(we use euclidean signature and complex coordinates 
$z =\frac{1}{\sqrt{2}} (x^0 +
ix^1)$, $\bar{z} =\frac{1}{\sqrt{2}} (x^0 - ix^1)$)
\bea
&&
2i\pab b \equiv \pab \left( e_{\ast}^{-\frac{i}{2} \phi}  \ast 
\pa e_{\ast}^{\frac{i}{2} \phi} 
- e_{\ast}^{\frac{i}{2} \phi}  \ast \pa e_{\ast}^{-\frac{i}{2} \phi}\right) 
~=~ i \g \sin_{\ast}{\phi}
\nonumber \\
&& 
2 \pab a \equiv \pab \left( e_{\ast}^{\frac{i}{2} \phi}  \ast 
\pa e_{\ast}^{-\frac{i}{2} \phi} 
+ e_{\ast}^{-\frac{i}{2} \phi}  \ast \pa e_{\ast}^{\frac{i}{2} \phi}\right) 
~=~ 0
\label{sg}
\eea
where $f \ast g = f \exp\left({\frac{\th}{2} (\overleftarrow{\pa}  
\overrightarrow{\bar{\pa}} -\overleftarrow{\bar{\pa}} \overrightarrow{\pa})}
\right)
g$.

The first equation contains the potential term which is the ``natural''
generalization of the ordinary sine potential, whereas the other one has 
the structure
of a conservation law and can be seen as imposing an extra condition
on the system. In the commutative limit, the first equation reduces to
the ordinary sine--Gordon equation, whereas the second one becomes trivial. 
The equations are in general complex and possess the $Z_2$ symmetry 
of the ordinary sine--Gordon ( invariance under $\phi \to -\phi$). 

The reason why integrability seems to require two equations of motions 
can be traced back to the general structure of unitary groups in NC geometry.
In the bicomplex approach the ordinary equations are obtained as 
zero curvature conditions for covariant derivatives defined in terms of
SU(2) gauge connections. If the same procedure is to be implemented
in the noncommutative case, the group SU(2), which is known to be not
closed in noncommutative geometry, has to be extended to a noncommutative
U(2) group and a NC U(1) factor enters necessarily into the game. 
The appearance of the second equation in (\ref{sg}) for our NC integrable 
version
of sine-Gordon is then a consequence of the fact that the fields develop
a nontrivial trace part. We note that the pattern of equations
we have found seems to be quite general
and unavoidable if integrability is of concern. In fact, the same 
has been found in Ref. \cite{CM} where a different but equivalent set of 
equations was proposed.  

The presence of two equations of motion is in principle very restrictive and
one may wonder whether the class of solutions is empty. To show that this
is not the case, in Ref. \cite{us} solitonic solutions were constructed 
perturbatively which reduce to the ordinary solitons when we take the
commutative limit. More generally, we observe that the second equation
in (\ref{sg}) is automatically satisfied by any chiral or antichiral 
function.
Therefore, we expect the class of solitonic solutions to be at least as
large as the ordinary one. In the general case, instead, 
we expect the class of dynamical
solutions to be smaller than the ordinary one because of the presence
of the nontrivial constraint. However, since the constraint equation is one 
order higher with respect to the dynamical equation, order by order in the 
$\theta$-expansion a solution always exists. This means that a 
Seiberg--Witten map between 
the NC and the ordinary model does not exist as a mapping between 
physical configurations, but it might be constructed as a mapping between 
equations of motion or conserved currents.  

The question which was left open in Ref. \cite{us} was the existence of an
action for the set of equations (\ref{sg}). In this paper we give an action
and discuss the relation of our model with the NC selfdual Yang--Mills theory
and the NC Thirring model. Moreover, we discuss some properties of the
corresponding S-matrix which, in spite of integrability, turns out to be
acausal and not factorized.

\vskip 15pt
\section{ Connection with NC selfdual Yang--Mills}
 
The (anti-)selfdual Yang--Mills equation is well-known to describe a
completely integrable classical system in four dimensions \cite{Yang}. 
In the ordinary case the equations of motion for many two dimensional 
integrable systems, including sine--Gordon, 
can be obtained through dimensional reduction of the (A)SDYM equations 
\cite{SDYMred}.

A convenient description of the (A)SDYM system is the so called 
$J$-formulation, given in terms of a $SL(N,C)$ 
matrix-valued $J$ field satisfying
\beq
\pa_{\bar{y}}\left(J^{-1} \pa_y J\right)+\pa_{\bar{z}}\left(J^{-1}\pa_z 
J\right)=0
\label{selfdual}
\eeq
where $y$, $\bar{y}$, $z$, $\bar{z}$ are complex variables 
treated as formally independent.

In the ordinary case, the sine-Gordon equation can be obtained from
(\ref{selfdual}) by taking $J$ in $SL(2, C)$ to be \cite{SDYMsine}
\beq
J=J(u,z,\bar{z})=e^{\frac{z}{2}\s_i}e^{\frac{i}{2}u\s_j}
e^{-\frac{\bar{z}}{2}\s_i}
\label{ansatz1}
\eeq
where $u=u(y,\bar y)$ depends on $y$ and $\bar y$ only and $\s_i$ are 
the Pauli matrices.

A noncommutative version of the (anti-)selfdual Yang--Mills 
system can be naturally obtained \cite{NCSDYM1}
by promoting the variables $y$, $\bar y$,
$z$ and $\bar z$ to be noncommutative thus extending the ordinary products in
(\ref{selfdual}) to $\ast$--products. In this case the $J$ field lives
in $GL(N,C)$. 

It has been shown \cite{NCSDYM3} that NC SDYM naturally emerges 
from open $N=2$ strings in a 
B-field background.
Moreover, in Refs. \cite{NCSDYM1, NCSDYM2, NCbur} examples of reductions to 
two-dimensional 
NC systems were given. It was also argued that  the NC deformation should 
preserve the integrability of the systems \cite{NCSDYM4}.  

We now show that our NC version of the sine-Gordon equations can be 
derived through dimensional reduction from the NC SDYM equations.
For this purpose we consider the NC version of equations (\ref{selfdual})
and choose $J_\ast$ in $GL(2,C)$ as
\beq
J_\ast=J_\ast(u,z,\bar{z})=e_\ast^{\frac{z}{2}\s_i}\ast 
e_\ast^{\frac{i}{2}u\s_j}\ast e_\ast^{-\frac{\bar{z}}{2}\s_i}
\label{ansatz2}
\eeq
This leads to the matrix equation
\beq
\pa_{\bar{y}}a~ I ~+~ i\left(\pa_{\bar{y}}b+\frac{1}{2} 
\sin_\ast{u} \right)\s_j=0
\label{sgmatrix}
\eeq
where $a$ and $b$ have been defined in (\ref{sg}).
Now, taking the trace we obtain $\pa_{\bar{y}} a = 0$  which is 
the constraint equation in (\ref{sg}). As a consequence, the term proportional
to $\s_j$ gives rise to the dynamical equation in (\ref{sg}) for the
particular choice $\g = -1$.
Therefore we have shown that the equations of motion of the NC version
of sine--Gordon proposed in Ref. \cite{us} can be obtained from a suitable 
reduction of the NC SDYM system as in the ordinary case. From this derivation 
the origin of the constraint appears even more clearly: 
it arises from setting to zero the trace part which the 
matrices in $GL(2,C)$ naturally develop under $\ast$--multiplication. 

Solving (\ref{sgmatrix}) for the particular choice $\s_j = \s_3$ we obtain the 
alternative set of equations
\bea
&& \pab \left( e_{\ast}^{-\frac{i}{2} \phi}  \ast \pa 
e_{\ast}^{\frac{i}{2} \phi}\right) ~=~\frac{i}{2} \g \sin_{\ast}{\phi} 
\label{sg2} \\
&& \pab \left(e_{\ast}^{\frac{i}{2} \phi}  \ast \pa 
e_{\ast}^{-\frac{i}{2} \phi} \right) ~=~ -\frac{i}{2} \g \sin_{\ast}{\phi}  
\nonumber
\eea
Order by order in the $\theta$-expansion the set of equations (\ref{sg}) 
and (\ref{sg2}) are equivalent. Therefore, the set (\ref{sg2}) is equally
suitable for the description of an integrable noncommutative generalization 
of sine-Gordon.

Since
our NC generalization of sine--Gordon is integrable, the present result
gives support to the arguments in favor of the integrability of NC SDYM 
system.

We note that our equations of motion, Wick rotated to Minkowski, can also be
obtained by suitable reduction of the $2+1$ integrable noncommutative model 
studied in Ref. \cite{olaf}.

\vskip 15pt
\section{The action} 
We are now interested in the possibility 
of determining an action for the scalar field $\phi$ satisfying the
system of eqs. (\ref{sg}). We are primarily motivated 
by the possibility to move on to a quantum description of the system. 

In general, it is not easy to find an action for the dynamical equation 
(the first eq. in (\ref{sg})) since $\phi$ is constrained by the second
one. One possibility could be to implement the constraint by the use
of a Lagrange multiplier.

We consider instead the equivalent set of equations (\ref{sg2}). 
We rewrite them in the form
\bea
&&\pab(g^{-1}\ast \pa g)=\frac{1}{4}\g\left(g^2-g^{-2}\right)\cr
&&\pab(g\ast\pa g^{-1})=-\frac{1}{4}\g\left(g^2-g^{-2}\right)
\label{sg3}
\ena
where we have defined $g\equiv e_\ast^{\frac{i}{2}\phi}$. Since $\phi$ is 
in general complex $g$ can be seen as an element of a noncommutative
complexified $U(1)$. The gauge group valued function 
$\bar g\equiv (g^\dagger)^{-1}= e_\ast^{\frac{i}{2}\phi^\dagger}$ 
is subject to the equations
\bea
&&\pab(\bar g \ast\pa\bar g^{-1})=-\frac{1}{4}\g \left(\bar g^2-\bar 
g^{-2}\right)\cr
&&\pab(\bar g^{-1}\ast\pa \bar g)=\frac{1}{4}\g \left(\bar g^2-\bar 
g^{-2}\right)
\label{sg4}
\ena
obtained by taking the h.c. of (\ref{sg3}).

In order to determine the action it is convenient to concentrate on the
first equation in (\ref{sg3}) and the second one in (\ref{sg4}) as the two
independent complex equations of motion which describe the dynamics of our 
system. 

We first note that the left-hand sides of equations (\ref{sg3}) and
(\ref{sg4}) have the chiral structure which is well known to correspond to a
NC version of the WZNW action \cite{NCWZ}. Therefore we are led to consider 
the action 
\beq
S[g,\bar g]=S[g]+S[\bar g] 
\label{action}
\eeq
where, introducing the homotopy path $\hat{g}(t)$ such that $\hat{g}(0)=1$,
$\hat{g}(1) = g$ ($t$ is a commuting parameter) we have defined
\bea
&S[g]& =  \int d^2z ~\left[ \pa  g \ast 
\pab g^{-1} + \int_0^1 dt ~ \hat{g}^{-1} \ast \pa_{t} \hat{g} \ast
[\hat{g}^{-1} \ast  \pa  \hat{g}, 
\hat{g}^{-1} \ast \pab  \hat{g} ]_{\ast} 
-\frac{\g}{4}(g^2+ g^{-2}-2) \right] \nonumber \\
&&~~~~~~~~~~
\label{actionWZNW}
\eea
and similarly for $S[\bar{g}]$. The first part of the action can be recognized 
as the NC generalization of a complexified $U(1)$ WZNW action \cite{NS}.

To prove that this generates the correct equations, we should take 
the variation with
respect to the $\phi$ field ($g = e_\ast^{\frac{i}{2}\phi}$)
and deal with complications which follow from
the fact that in the NC case the variation of an exponential is not
proportional to the exponential itself.
However, since the variation $\d \phi$ is arbitrary, 
we can forget about its $\theta$ dependence and write $\frac{i}{2}\d \phi 
= g^{-1} \d g $, trading the variation 
with respect to $\phi$ with the variation
with respect to $g$. Analogously, the variation with respect to 
$\phi^{\dagger}$ can be traded with the variation with respect to $\bar{g}$.

It is then a simple calculation to show that
\beq
\d S[g] = \int d^2z ~2g^{-1} \d g \left[ \pab \left( g^{-1} 
\ast \pa g \right) ~-~ \frac{i}{2} \g \sin_{\ast}{\phi}
\right] 
\eeq
from which we obtain the first equation in (\ref{sg3}). Treating $\bar g$
as an independent variable an analogous derivation gives the second equation 
in (\ref{sg4}) from $S[\bar g]$.

We note that, when $\phi$ is real, $g=\bar g$ and the action 
(\ref{action}) reduces to $S_{\rm real}[g]=2S_{WZW}[g]-\g(\cos_\ast\phi -1)$. 
In general, since the two equations (\ref{sg}) are complex it
would be inconsistent to restrict ourselves to real solutions. However, it is
a matter of fact that the equations of motion become real when the field
is real.
Perturbatively in $\theta$ this can be proved order by order 
by direct inspection of the equations in Ref. \cite{us}. In particular, at
a given order one can show that the imaginary part of the equations
vanishes when the constraint and the equations of motion at
lower orders are satisfied. 

It would be interesting to obtain the action (\ref{action}) from the 
dimensional reduction of the 4d SDYM action by generalizing to the NC case
the procedure used in Ref. \cite{SDYMsine}.

\vskip 15pt
\section{Connection with NC Thirring model}

In the ordinary case the equivalence between Thirring and  
sine--Gordon models \cite{coleman} can be proven at the level of functional
integrals by implementing the bosonization prescription 
\cite{witten, pw, divecchia} on the fermions.
The same procedure has been worked out in NC geometry \cite{MS,NOS}.
Starting from the NC version of Thirring described by 
\beq
S_T = \int d^2x \left[ \bar{\psi} i \g^\mu \pa_\mu \psi + m \bar{\psi} \psi
- \frac{\l}{2}(\bar{\psi} \ast \g^\mu \psi)  (\bar{\psi} \ast \g_\mu \psi) 
\right]
\label{Thirring}
\eeq
the bosonization prescription gives rise to the action for the bosonized 
NC massive Thirring model which turns out to be a 
NC WZNW action supplemented by a cosine potential term for the NC U(1)
group valued field which enters the bosonization of the fermionic currents.
In particular, in the most recent paper in Ref. \cite{MS} it has been
shown that working in Euclidean space the massless Thirring action 
corresponds to the sum of two WZNW actions once a suitable choice for the 
regularization parameter is made. Moreover, in Ref. \cite{NOS} it was proven 
that the bosonization of the mass term in (\ref{Thirring}) gives rise to
a cosine potential for the scalar field with coupling constant proportional
to $m$.

The main observation is that our action (\ref{action}) is the sum of two
NC WZNW actions plus cosine potential terms for the pair of 
$U(1)_C$ group valued fields $g$ and $\bar g$, considered as independent.
Therefore, our action can be interpreted as coming from the bosonization
of the massive NC Thirring model, in agreement with the results 
in Refs. \cite{MS,NOS}.

We have shown that even in the NC case
the sine--Gordon field can be interpreted
as the scalar field which enters the bosonization of the Thirring model,
so proving that the equivalence between Thirring and 
sine--Gordon can be maintained in NC generalizations of these models. 
Moreover, the classical integrability of our NC version of 
sine--Gordon proven in Ref. \cite{us} should automatically guarantee the 
integrability of the NC Thirring model.  

In the particular case of zero coupling ($\g = 0$), the equations 
(\ref{action})
and (\ref{sg2}) correspond to the action and the equations of motion
for a NC U(1) WZNW model \cite{NCWZ}, respectively. Again, we can use 
the results
of Ref. \cite{us} to prove the classical integrability of the NC U(1) WZNW
model and construct explicitly its conserved currents.

\section{Properties of the S-matrix}

It is well known that in integrable commutative field theories there is no 
particle production and the S-matrix factorizes. In the noncommutative case
properties of the S-matrix have been investigated for two
specific models: The $\l \Phi^4$ theory in two dimensions \cite{seiberg} and
the nonintegrable ``natural'' NC generalization the the sine--Gordon model
\cite{CM}. In the first reference a very pathological acausal behavior was 
observed due to the space and time noncommutativity. For an incoming wave
packet the scattering produces an advanced wave which arrives at the origin
before the incoming wave. In the second model investigated it was found
that particle production occurs. The tree level $2 \to 4$ amplitude does not 
vanish.   

It might be hoped that classical integrability would alleviate these 
pathologies. In the NC integrable sine-Gordon case, since we have an action,
it is possible to investigate these issues. As described below we have 
computed the scattering amplitude for the $2 \to 2 $ process and found that
the acausality of Ref. \cite{seiberg} is not cured by integrability.  
We have also computed the production amplitudes for the processes $2 \to 3$ 
and $2 \to 4$ and found that they don't vanish.

We started from our action (\ref{actionWZNW}) rewritten in terms of Minkowski 
space coordinates $x^0,x^1$ and real fields 
($g = e_{\ast}^{\frac{i}{2} \phi}$, $\hat{g}(t) = 
e_{\ast}^{\frac{i}{2} t \phi}$ with $\phi$ real)

\bea
&&S[g]=-\frac{1}{2}\int d^2 x~ g^{-1}\ast \pa^\mu g \ast
g^{-1}\ast\pa_\mu g -
\frac{1}{3}\int d^3 x ~\e^{\m\n\rho}\hat{g}^{-1}\ast \pa_\mu \hat{g} \ast 
\hat{g}^{-1}\ast\pa_\n \hat{g} \ast \hat{g}^{-1}\ast\pa_\rho \hat{g} \cr
&&~~~~~+\frac{\g}{4}\int d^2 x (g^2+g^{-2}-2) 
\ena
where $f \ast g = f e^{\frac{i}{2} \th \e^{\mu\nu} \overleftarrow{\pa}_\mu
\overrightarrow{\pa}_\nu} g$, and we derived the following Feynman's rules

\begin{itemize}
\item The propagator
\beq
G(q)=\frac{4i}{q^2-2\g}
\eeq

\item
The vertices 
\bea
&& v_3(k_1,\dots,k_3)= \frac{2}{2^3 \cdot 3!}
\e^{\m\n}k_{1\m}k_{2\n}F(k_1,\dots,k_3) \nonumber \\
&&v_4(k_1,\dots,k_4)= i \left(-\frac{1}{2^4 \cdot 4!}
\left(k_1^2+3k_1\cdot k_3\right)+\frac{\g}{2 \cdot 4!}\right)
F(k_1,\dots,k_4) \nonumber \\
&& v_5(k_1,\dots,k_5)= -\frac{2\e^{\m\n}}{2^5 \cdot 5!}
\left(k_{1\m}k_{2\n}-k_{1\m}k_{3\n}+2k_{1\m}k_{4\n}\right)F(k_1,\dots,k_5)
\nonumber \\
&&  v_6(k_1,\dots,k_6)= i\left[\frac{1}{2^6 \cdot 6!}
\left(k_{1}^2+5k_{1}\cdot k_{3}-5k_{1} \cdot k_{4}+5k_{1} \cdot k_{5}\right)
- \frac{\g}{2 \cdot 6!}\right] F(k_1,\dots,k_6)
\nonumber \\
&&~~~~~~~~~
\label{vertices}
\ena
where
\beq
F(k_1,\dots,k_n)=\exp\left(-\frac{i}{2}\sum_{i<j}k_i\times k_j\right)
\eeq
is the phase factor coming from the $\ast$-products in the action 
(we have indicated $a\times b= \theta \e^{\m\n} a_\m b_\n$), $k_i$ are
all incoming momenta and we used momentum conservation.
\end{itemize}

At tree level the $2 \to 2$ process is described by the diagrams with
the topologies in Fig. 1.

\vskip 18pt
\noindent
%---------- FIGURE TOP ------------
\begin{minipage}{\textwidth}
\begin{center}
\includegraphics[width=0.60\textwidth]{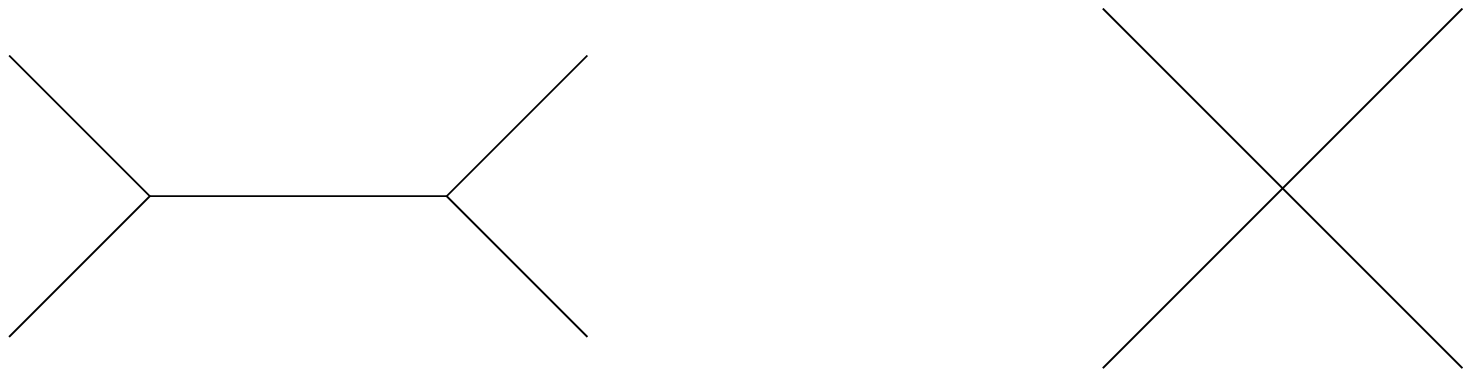}
\end{center}
\begin{center}
{\small{Figure 1:
Tree level $2 \to 2$ amplitude}}
\end{center}
\end{minipage}
%---------- FIGURE END ------------

\vskip 20pt

Including contributions from the various channels and using the three point
and four point vertices of eqs. (\ref{vertices}) we obtained for the
scattering amplitude the expression  
\beq
-\frac{i}{2} E^2 p^2 \left( \frac{1}{2E^2 - \g} - \frac{1}{2p^2 + \g}
\right) \sin^2{(pE \th)} ~+~ i \frac{\g}{2} \cos^2{(pE \th)}
\label{final}
\eeq
where $p$ is the center of mass momentum and $E = \sqrt{p^2 + 2\g}$.

For comparison with Ref. \cite{seiberg} this should be multiplied by an
incoming wave packet  
\beq
\Phi_{in}(p) \sim \left( e^{-\frac{(p -p_0)^2}{\l}}+ 
e^{-\frac{(p +p_0)^2}{\l}}\right)
\eeq
and Fourier transformed with $e^{ipx}$.
We have not attempted to carry out the Fourier transform integration.
However, we note that for $p_0$ very large $E$ and $p$ are concentrated
around large values and the scattering amplitude assumes the form 
\beq
i\frac{\g}{4} \sin^2{(pE \th)} ~+~ i \frac{\g}{2} \cos^2{(pE \th)}
\eeq
which is equivalent to the result in Ref. \cite{seiberg}, leading to
the same acausal pathology \footnote{It is somewhat tantalizing that 
a change in the relative coefficient between the two terms 
would lead to a removal of the trigonometric factors which are responsible
for the acausal behavior.}. 

We describe now the computation of the production amplitudes $2 \to 3$ and 
$2 \to 4$. At tree level the contributions are drawn in Figures 2 and 3, 
respectively.

\vskip 18pt
\noindent
%---------- FIGURE TOP ------------
\begin{minipage}{\textwidth}
\begin{center}
\includegraphics[width=0.60\textwidth]{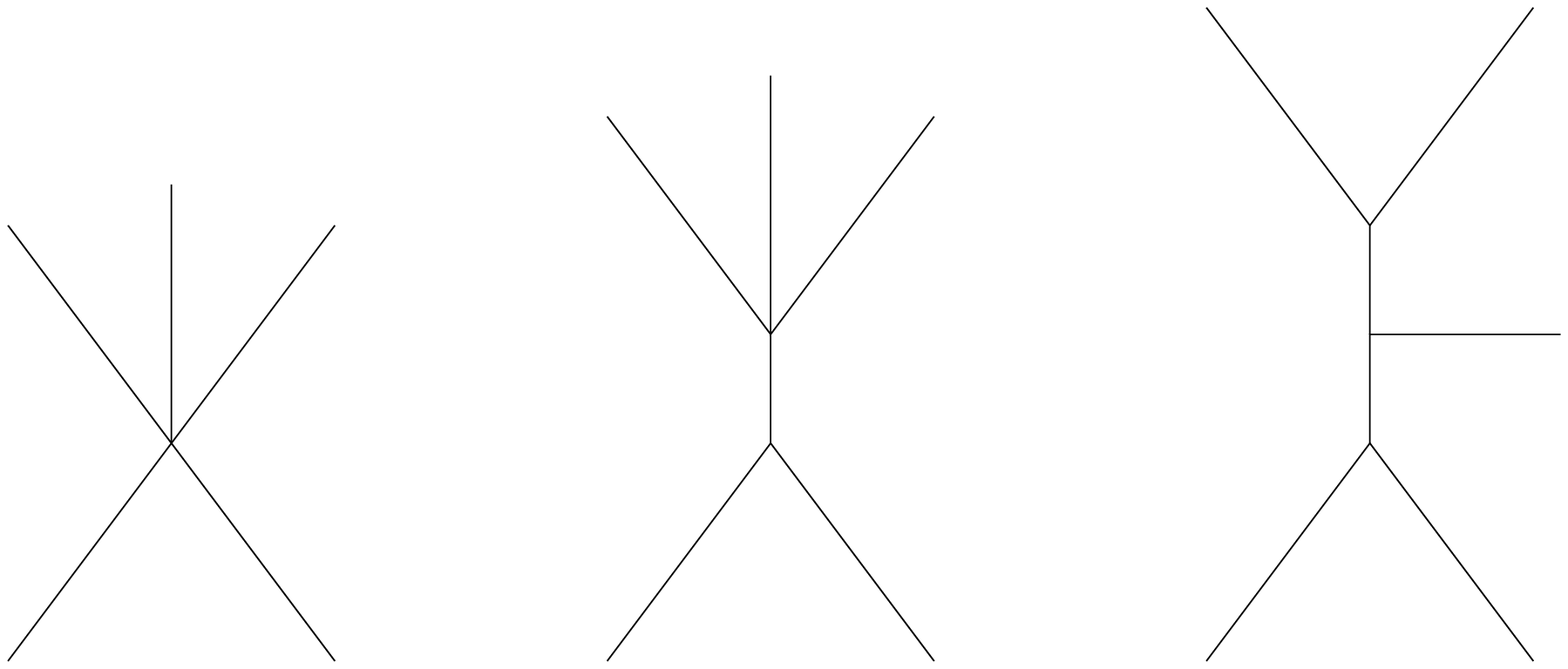}
\end{center}
\begin{center}
{\small{Figure 2:
Tree level $2 \to 3$ amplitude}}
\end{center}
\end{minipage}
%---------- FIGURE END ------------

\vskip 18pt
\noindent
%---------- FIGURE TOP ------------
\begin{minipage}{\textwidth}
\begin{center}
\includegraphics[width=0.60\textwidth]{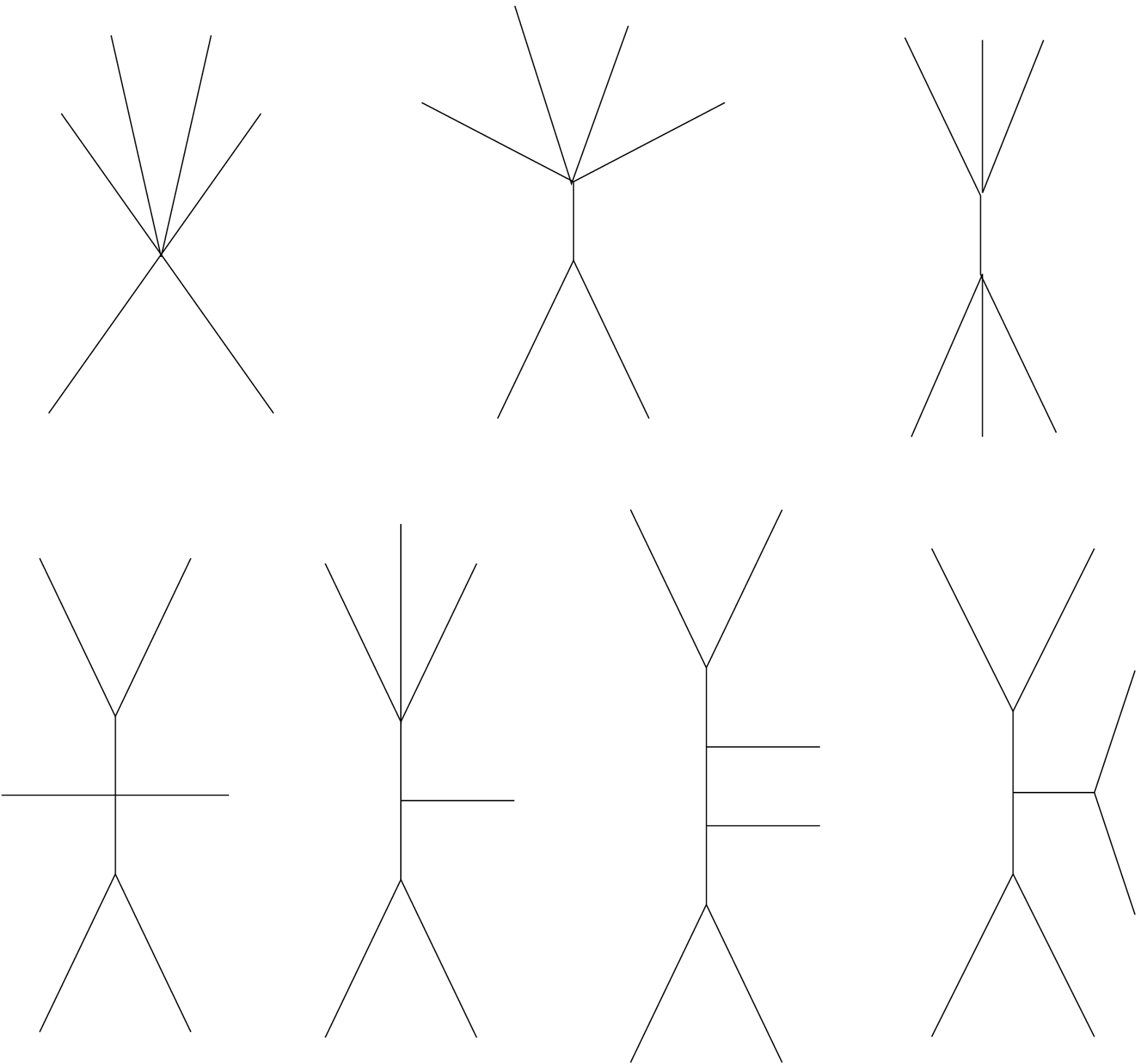}
\end{center}
\begin{center}
{\small{Figure 3:
Tree level $2 \to 4$ amplitude}}
\end{center}
\end{minipage}
%---------- FIGURE END ------------

\vskip 20pt

For any topology the different possible channels must be taken into account. 
This, as well as the complicated expressions for the vertices, has led us
to use an algebraic manipulation program computer.
We used {\em Mathematica}${}^{\copyright}$ to symmetrize completely 
the vertices 
(\ref{vertices}). This allows to take automatically into account the 
different diagrams obtained by exchanging momenta entering a given vertex. 
The contribution from each diagram was obtained as a product of 
the combinatorial factor, the relevant vertices and propagators.  
Due to the length of the program it was impossible to handle the calculation
in a complete analytic way. Instead, the program was run with assigned 
values of the momenta and arbitrary  $\theta$ and $\g$. For both the  
$2\rightarrow 3$ and $2\rightarrow 4$ processes the result is nonvanishing.
As a check of our calculation we mention that the production
amplitudes vanish  when we set $\th = 0$, for any value of the coupling and 
the momenta.

This investigation has been carried on in the particular case 
$\phi = \phi^\dag$. However, it can be easily proved that turning on a nontrivial 
imaginary part for the $\phi$ field does not cure the previous pathologies.

\vskip 15pt
\section{Conclusions}

In this paper we have investigated some properties of the integrable NC
sine--Gordon system proposed in Ref. \cite{us}. We succeeded in constructing
an action which turned out to be a WZNW action for a noncommutative, 
complexified $U(1)$ augmented by a cosine potential. We have shown that 
even in the NC case there is a duality relation between our integrable
NC sine--Gordon model and the NC Thirring. 

NC WZNW models have been shown to be one--loop renormalizable \cite{LFI}. 
This suggests that the NC sine--Gordon model proposed
in Ref. \cite{us} is not only integrable but it might lead to a well-defined
quantized model, giving support to the existence of a possible relation 
between integrability and renormalizability. 

Armed with our action we investigated some properties of the S--matrix
for elementary excitations.
However, in contradistinction to the commutative case, the S--matrix turned out
to be acausal and nonfactorizable \footnote{Other problems of the S-matrix have
been discussed in Ref. \cite{gmbk}.}. The reason for the acausal behavior 
has been
discussed in Ref. \cite{seiberg}~ where it was pointed out that 
noncommutativity induces a backward-in-time effect because of the presence
of certain phase factors. It appears that in our case this effect is still
present in spite of integrability. 

It is not clear why the presence of an infinite number of local conserved
currents (local in the sense that they are not expressed as integrals of 
certain densities) does not guarantee factorization and absence of 
production in the
S-matrix as it does in the commutative case. The standard proofs of 
factorization use, among other assumptions, the mutual commutativity of the
charges - a property we have not been able to check so far because of the
complicated nature of the currents. But even if the charges were to commute
the possibility of defining them as powers of momenta, as required in the
proofs, could be spoiled by
acausal effects which prevent a clear distinction between incoming and
outgoing particles. 
Indeed, this may indicate some fundamental inconsistency of the
model as a field theory describing scattering in 1+1 dimensional Minkowski
space (simply going to Euclidean signature does not change the
factorizability  properties of the S-matrix). However, it is conceivable
that in  Euclidean space the model could be used to
describe some statistical mechanics system and this possibility might be
worth investigating.

In a series of papers \cite{liao} a different approach to quantum NC 
theories has 
been proposed when the time variable is not commuting. 
In particular, the way one computes Green's functions is different there,
leading to a modified definition of the S-matrix. It would be 
interesting to redo our calculations in that approach to see whether a 
well-defined factorized S-matrix for our model can be constructed. 
In this context it would be also interesting to investigate the scattering 
of solitons present in our model \cite{us}. To this end, since our
model is a reduction of the $2+1$ integrable model studied in \cite{olaf2},
it might be possible to exploit the results of those papers concerning 
multi--solitons and their scattering to investigate the same issues in 
our case.    
 
The model we have proposed in the present paper 
describes the propagation of two interacting scalar fields $Re \phi$, 
$Im \phi$ . The particular form of the interaction follows from the
choice of the $U(2)$ matrices made in \cite{us} for the
bicomplex formulation or, equivalently, from the particular ansatz
(\ref{ansatz2}) in the reduction from NC SDYM (in the commutative limit
and for $Im\phi =0$ we are back to the ordinary sine--Gordon). 
In the commutative case the ansatz (\ref{ansatz1}) depends on a single
real field and, independently of the choice of the Pauli matrices
in the exponentials, we obtain the same equations of motion.
In the NC case the lack of decoupling of the U(1) subgroup requires
the introduction of two fields. This implies that in general we can
make an ansatz $J_\ast = e_{\ast}^{\frac{z}{2} \s_i} g(y, \bar{y})
e_{\ast}^{-\frac{\bar{z}}{2} \s_i}$
where $g$ is a group valued field which depends on two scalar fields in
such a way as to reduce to the ordinary ansatz (\ref{ansatz1}) in the 
commutative limit and for a suitable identification of the two fields.
In principle there are different choices for $g$ as a function of the two
fields satisfying this requirement. Different choices may be 
inequivalent and describe different but still integrable dynamics for the
two fields. Therefore, an interesting question  is
whether an ansatz slightly different from (\ref{ansatz2}) exists which 
would give 
rise to an integrable system described by a consistent, factorized
S-matrix.
We might expect that if such an ansatz exists it should introduce a
{\em different} interaction between the two fields and this might cure the
pathological behavior of the
present scattering matrix. If such a possibility exists it would be
interesting to compare the two different reductions to understand in the
NC case what really drives the system to be integrable in the sense of
having a well--defined, factorized S--matrix since the existence of an
infinite number of local conservation laws does not appear to be
sufficient. These issues will be discussed in a forthcoming paper
\cite{LMPPT}.

\medskip

\section*{Acknowledgements}
\noindent  We acknowledge useful discussions with O. Lechtenfeld and
A. Popov. L.T. would like to thank A. Marzuoli, V. Korepin and M. Ro\v cek
for useful discussions and the YITP at Stony Brook for the kind hospitality. 
M.T.G. would like to thank the Physics Department of Milano-Bicocca University 
for hospitality during the period when some of this work was done.
This work has been supported in 
part by INFN, MURST and the European Commission RTN program
HPRN--CT--2000--00131, in which S.P. is associated to
the University of Padova. The work of M.T.G. is supported by NSF Grant 
No. PHY-0070475 and by NSERC Grant No. 204540.

\end{document}